\documentclass[letterpaper]{sig-alternate-2013}

\usepackage{amsmath,amscd,amsbsy,amssymb,latexsym,url,bm}
\usepackage{epsfig,epstopdf,graphicx,subfigure}
\usepackage{enumitem,balance,mathtools}
\usepackage{wrapfig}
\usepackage{mathrsfs, euscript}
\usepackage[usenames]{color}
\usepackage{multicol,multirow}
\usepackage{algorithm}
\usepackage{algorithmic}
\usepackage{caption}

\newfont{\mycrnotice}{ptmr8t at 7pt}
\newfont{\myconfname}{ptmri8t at 7pt}
\let\crnotice\mycrnotice%
\let\confname\myconfname%

\permission{~}
\conferenceinfo{WSDM 2016,}{February 22-25, 2016, San Francisco, CA, USA.}
\copyrightetc{arXiv version}
\crdata{}

\usepackage[margin=0.755in]{geometry}
\begin{document}

\title{Feedback Control of Real-Time Display Advertising}

\numberofauthors{1}
\author{
\alignauthor Weinan Zhang$^1$, Yifei Rong$^{2,1}$, Jun Wang$^1$, Tianchi Zhu$^3$, Xiaofan Wang$^4$\\
       \affaddr{$^1$University College London, $^2$YOYI Inc., $^3$Big Tree Times Co., $^4$Shanghai Jiao Tong University}\\
       \email{$^1$\{w.zhang, j.wang\}@cs.ucl.ac.uk, $^2$yifei.rong@yoyi.com.cn}\\
       \email{$^3$tc@bigtree.mobi, $^4$xfwang@sjtu.edu.cn}\\
}


\maketitle

\begin{abstract}
  Real-Time Bidding (RTB) is revolutionising display advertising by facilitating per-impression auctions to buy ad impressions as they are being generated. Being able to use impression-level data, such as user cookies, encourages user behaviour targeting, and hence has significantly improved the effectiveness of ad campaigns.  However, a fundamental drawback of RTB is its instability because the bid decision is made per impression and there are enormous fluctuations in campaigns' key performance indicators (KPIs).  As such, advertisers face great difficulty in controlling their campaign performance against the associated costs. In this paper, we propose a feedback control mechanism for RTB which helps advertisers dynamically adjust the bids to effectively control the KPIs, e.g., the auction winning ratio and the effective cost per click. We further formulate an optimisation framework to show that the proposed feedback control mechanism also has the ability of optimising campaign performance. By settling the effective cost per click at an optimal reference value, the number of campaign's ad clicks can be maximised with the budget constraint. Our empirical study based on real-world data verifies the effectiveness and robustness of our RTB control system in various situations. The proposed feedback control mechanism has also been deployed on a commercial RTB platform and the online test has shown its success in generating controllable advertising performance.  \end{abstract}


\section{Introduction}
Emerged in 2009, Real-Time Bidding (RTB) has become a new paradigm in display advertising \cite{muthukrishnan2009ad,google2011arrival}.~Different from the conventional human negotiation or pre-setting a fixed price for impressions, RTB creates an impression-level auction and enables advertisers to bid for individual impression through computer algorithms served by demand-side platforms (DSPs) \cite{zhang2014optimal}. The bid decision could depend on the evaluation of both the utility (e.g., the likelihood and economic value of an impression for generating click or conversion) and the cost (e.g., the actual paid price) of each ad impression. More importantly, real-time information such as the specific user demographics, interest segments and various context information is leveraged to help the bidding algorithms evaluate each ad impression. With the real-time decision making mechanism, it is reported that RTB yields significantly higher return-on-investment (ROI) than other online advertising forms \cite{yuan2013real}.

\begin{figure}
 \centering
  \subfigure{
   \includegraphics[height=1.1in]{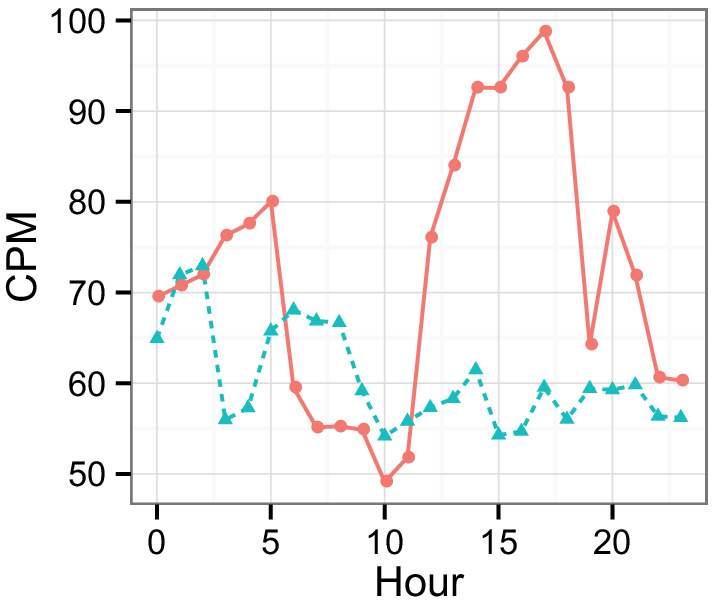}}
  \subfigure{
   \includegraphics[height=1.1in]{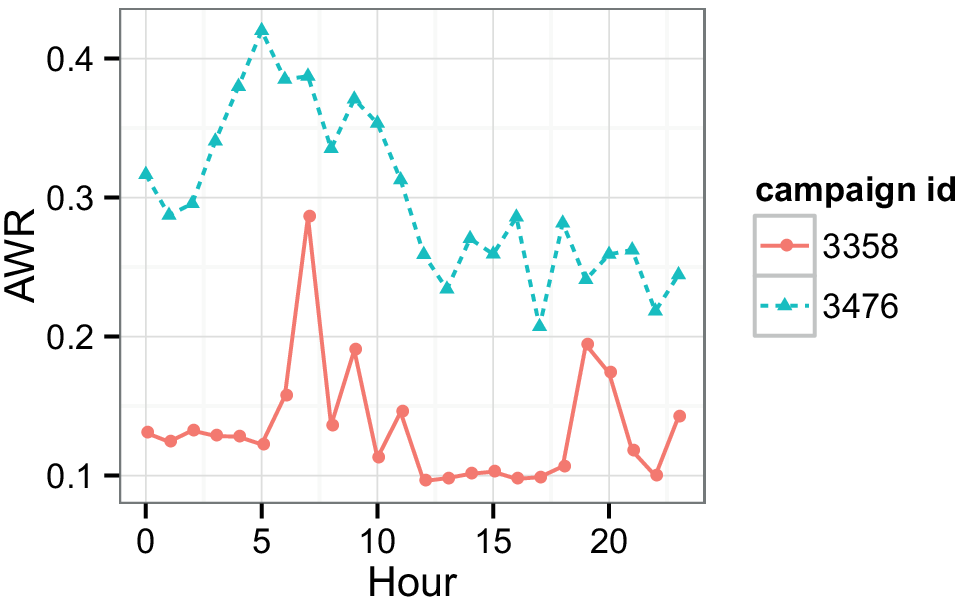}}
   \subfigure{
   \includegraphics[height=1.1in]{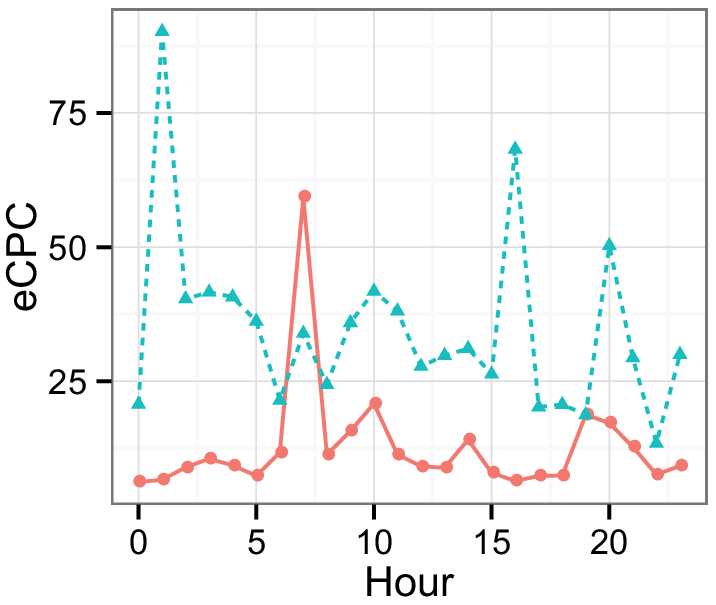}}
  \subfigure{
   \includegraphics[height=1.1in]{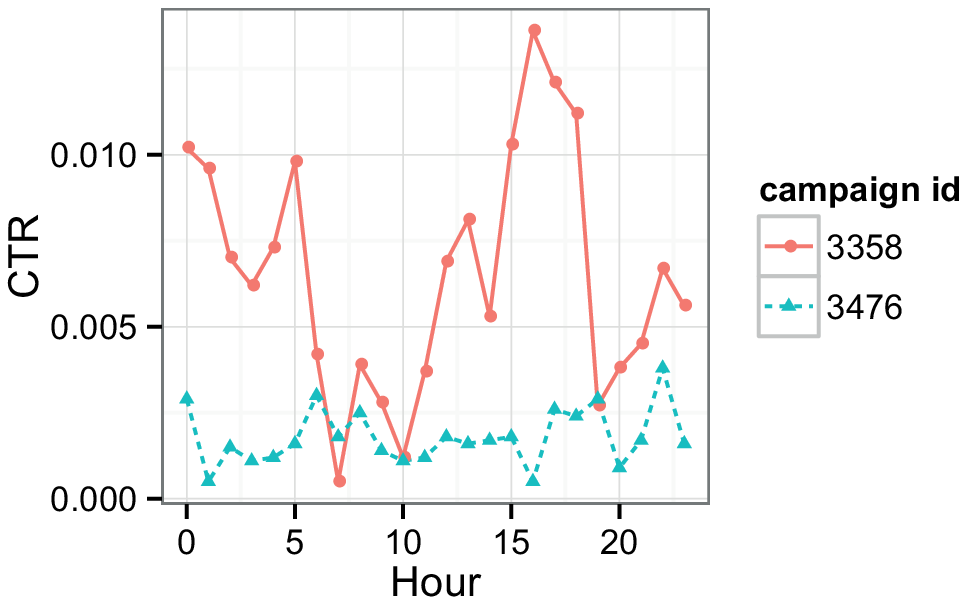}}
 \caption{The instability of CPM (cost per mille), AWR (auction winning ratio), eCPC (effective cost per click), and CTR (click-through rate) for two sample campaigns without a controller. Dataset: iPinYou.}
 \label{fig:instability}
\end{figure}

Despite the ability of delivering performance-driven advertising, RTB, unfortunately, results in high volatilities, measured by major Key Performance Indicators (KPIs), such as CPM (cost per mille), AWR (auction winning ratio), eCPC (effective cost per click) and CTR (click-through rate). To illustrate this, Figure~\ref{fig:instability} plots the four major KPIs over time for two example campaigns in a real-world RTB dataset. All four KPIs fluctuate heavily across the time under a widely-used bidding strategy \cite{perlich2012bid}.  Such instability causes advertisers ample difficulty in optimising and controlling the KPIs against their cost.

In this paper, we propose to employ feedback control theory \cite{aastrom2014control} to solve the instability problem in RTB. Feedback controllers are widely used in various applications for maintaining dynamically changing variables at the predefined reference values. The application scenarios range from the plane direction control \cite{nelson1998flight} to the robot artificial intelligence \cite{peterson1999decision}. In our RTB scenario, the specific KPI value, depending on the requirements from the advertisers, is regarded as the variable we want to control with a pre-specified reference value. Our study focuses on two use cases. (i) For performance-driven advertising, we concern with the feedback control of the average cost on acquiring a click, measured by effective cost per click (eCPC). (ii) For branding based advertising, to ensure a certain high exposure of a campaign, we focus on the control of the ratio of winning the auctions for the targeted impressions, measured by auction winning ratio (AWR).  More specifically, we take each of them as the control input signal and consider the gain (the adjustment value) of bid price as the control output signal for each incoming ad display opportunity (the bid request). We develop two controllers to test: the widely used proportional-integral-derivative (PID) controller \cite{bennett1993development} and the waterlevel-based (WL) controller \cite{dezotell1936water}. We conduct large-scale experiments to test the feedback control performance with different settings of reference value and reference dynamics.
Through the empirical study, we find that the PID and WL controllers are capable of controlling eCPC and AWR, while PID further provides a better control accuracy and robustness than WL.

Furthermore, we investigate whether the proposed feedback control can be employed for controllable bid optimisation. It is common that the performance of an ad campaign (e.g., eCPC) varies from different channels (e.g., ad exchanges, user geographic regions and PC/mobile devices) \cite{zhang2014real}. If one can reallocate some budget from less cost-effective channels to more cost-effective ones, the campaign-level performance would improve \cite{zhang2012joint}. In this paper, we formulate the multi-channel bid optimisation problem and propose a model to calculate the optimal reference eCPC for each channel. Our experiments show that the campaign-level click number and eCPC achieve significant improvements with the same budget.

Moreover, the proposed feedback control mechanism has been implemented and integrated in a commercial DSP. The conducted live test shows that in a real and noisy setting the proposed feedback mechanism has the ability to produce controllable advertising performance.

To sum up, the contributions of our work are as follows. (i) We study the instability problem in RTB and investigate its solution by leveraging the feedback control mechanism. (ii) Comprehensive offline and online experiments show that PID controller is better than other alternatives and finds the optimal way to settle the variable in almost all studied cases. (iii) We further discover that feedback controllers are of great potential to perform bid optimisation through settling the eCPC at the reference value calculated by our proposed mathematical click maximisation framework.

The rest of this paper is organised as follows. Section \ref{sec:preliminaries} provides preliminaries for RTB and feedback control. Our solution is formally presented in Section \ref{sec:system}. The empirical study is reported in Section \ref{sec:exp}, while the online deployment and live test are given in Section \ref{sec:online}. Section \ref{sec:related-work} discusses the related work and we finally conclude this paper in Section \ref{sec:conclusion}.

\section{Preliminaries} \label{sec:preliminaries} To make the paper self-contained, in this section, we take a brief review on the RTB eco-system, bidding strategies, and some basics of feedback control theory.

\subsection{RTB Flow Steps}
The interaction process among the main components of the RTB eco-system is summarised into the following steps: (0) when a user visits an ad-supported site (e.g., web pages, streaming videos and mobile apps), each ad placement will trigger a call for ad (ad request) to the ad exchange.  (1) The ad exchange sends the bid requests for this particular ad impression to each advertiser's DSP bidding agent, along with the available information such as the user and context information.  (2) With the information of the bid request and each of its qualified ads, the bidding agent calculates a bid price. Then the bid response (ad, bid price) is sent back to the exchange.  (3) Having received the bid responses from the advertisers, the ad exchange hosts an auction and picks the ad with the highest bid as the auction winner.  (4) Then the winner is notified of the auction winning from the ad exchange.  (5) Finally, the winner's ad will be shown to the visitor along with the regular content of the publisher's site.  It is commonly known that a long time page-loading would greatly reduce users' satisfactory \cite{muthukrishnan2009ad}.  Thus, advertiser bidding agents are usually required to return a bid in a very short time frame (e.g., 100 ms).  (6) The user's feedback (e.g., click and conversion) on the displayed ad is tracked and finally sent back to the winner advertiser. For a detailed discussion about RTB eco-systems, we refer to \cite{zhang2014real, yuan2013real}.  The above interaction steps have the corresponding positions in Figure~\ref{fig:control-in-rtb}, as we will discuss later.


\subsection{Bidding Strategies}
A basic problem for DSP bidding agents is to figure out how much to bid for an incoming bid request. The bid decision depends on two factors for each ad impression: the utility (e.g., CTR, expected revenue) and cost (i.e., expected charged price)~\cite{zhang2014optimal}. In a widely adopted bidding strategy \cite{perlich2012bid}, the utility is evaluated by CTR estimation while the base bid price is tuned based on the bid landscape \cite{cui2011bid} for the cost evaluation. The generalised bidding strategy in \cite{perlich2012bid} is
\begin{align}
b(t) = b_0 \frac{\theta_t}{\theta_0}, \label{eq:bid-lin}
\end{align}
where $\theta_t$ is the estimated CTR for the bid request at moment $t$; $\theta_0$ is the average CTR under a target condition (e.g., a user interest segment); and $b_0$ is the tuned base bid price for the target condition.  In this work, we adopt this widely used bidding strategy and adopt a logistic CTR estimator \cite{richardson2007predicting}.


\subsection{Feedback Control Theory}

Feedback control theory deals with the reaction and control of dynamic systems from feedback and outside noise \cite{aastrom2014control}. 
The usual objective of feedback control theory is to control a dynamic system so that the system output follows a desired control signal, called the reference, which may be a fixed or changing value. To attain this objective, a controller is designed to monitors the output and compares it with the reference. The difference between actual and desired output, called the error factor, is applied as feedback from the dynamic system to the control system. With the specific control function, the controller outputs the control signal, which is then transformed by the actuator into the system input signal sent back to the dynamic system. These processes \text{form} a feedback control loop between the dynamic system and the controller. Control techniques are widely used in various engineering applications for maintaining some signals at the predefined or changing reference values, such as plane navigation \cite{nelson1998flight} and water distribution control \cite{dezotell1936water}.

\section{RTB Feedback Control System} \label{sec:system}

\begin{figure}
 \centering
 \includegraphics[width=\columnwidth]{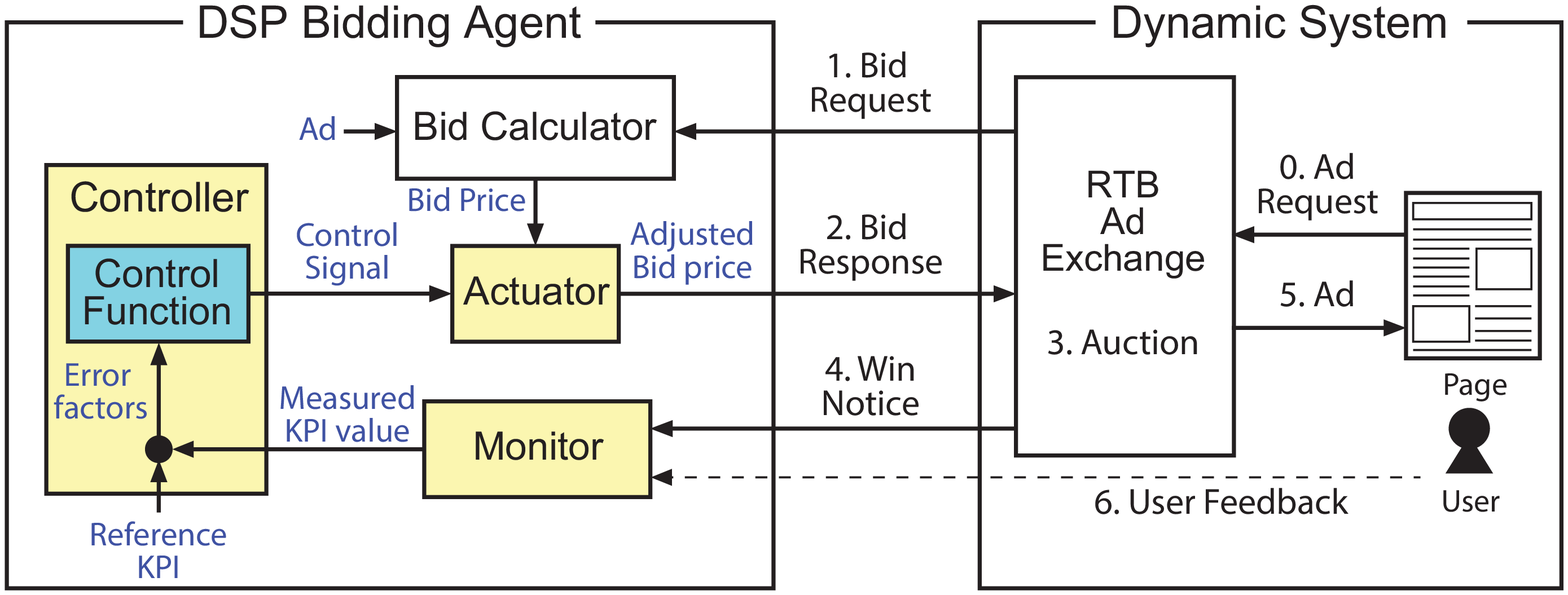}
 \caption{Feedback controller integrated in the RTB system.}
 \label{fig:control-in-rtb}
\end{figure}

Figure~\ref{fig:control-in-rtb} presents the diagram of the proposed RTB feedback control system. The traditional bidding strategy is represented as the \emph{bid calculator} module in the DSP bidding agent. The controller plays as a role which adjusts the bid price from the bid calculator.

Specifically, the \emph{monitor} receives the auction win notice from the ad exchange and the user click feedback from the ad tracking system, which as a whole we regard as the dynamic system. Then the current KPI values, such as AWR and eCPC can be calculated. If the task is to control the eCPC with the reference value, the error factor between the reference eCPC and the measured eCPC is calculated then sent into the control function. The output control signal is sent to the \emph{actuator}, which uses the control signal to adjust the original bid price from the bid calculator. The adjusted bid price is packaged with the qualified ad into the bid response and sent back to the ad exchange for auction.

\subsection{Actuator}\label{sec:actuator}
For the bid request at the moment $t$, the actuator takes into the current control signal $\phi(t)$ to adjust the bid price from $b(t)$ (Eq.~(\ref{eq:bid-lin})) to a new value $b_a(t)$. In our model, the control signal, which will be mathematically defined in the next subsections, is a gain on the bid price. Generally, when the control signal $\phi(t)$ is zero, there should be no bid adjustment. There could be different actuator models, and in our work we choose to use
\begin{align}
b_a(t) = b(t) \exp\{\phi(t)\}, \label{eq:exp-actuator}
\end{align}
where the model satisfies $b_a(t) = b(t)$ when $\phi(t) = 0$. Other models such as the linear model $b_a(t) \equiv b(t) (1+\phi(t))$ are also investigated in our study but it performs poorly in the situations when a big negative control signal is sent to the actuator, where the linear actuator will usually respond a negative or a zero bid, which is meaningless in our scenario. By contrast, the exponential model is a suitable solution to addressing the above drawback because it naturally avoids generating a negative bid. In the later empirical study we mainly report the analysis based on the exponential-form actuator model.

\subsection{PID Controller}\label{sec:pid}
The first controller we investigate is the classic PID controller \cite{bennett1993development}. As its name implies, a PID controller produces the control signal from a linear combination of the proportional factor, the integral factor and the derivative factor based on the error factor:
\begin{align}
e(t_k) &= x_r - x(t_k) \label{eq:error-factor},  \\
\phi(t_{k+1}) &\leftarrow \lambda_P e(t_k) + \lambda_I \sum_{j=1}^{k} e(t_j) \triangle t_j + \lambda_D \frac{\triangle e(t_k)}{\triangle t_k},  \label{eq:pid}
\end{align}
where the error factor $e(t_k)$ is the reference value $x_r$ minus the current controlled variable value $x(t_k)$, the update time interval is given as $\triangle t_j = t_j - t_{j-1}$, the change of error factors is $\triangle e(t_k) = e(t_k) - e(t_{k-1})$, and $\lambda_P$, $\lambda_I$, $\lambda_D$ are the weight parameters for each control factor. Note that here the control factors are all in discrete time ($t_1, t_2, \ldots$) because bidding events are discrete and it is practical to periodically update the control factors.~All control factors ($\phi(t), e(t_k), \lambda_P, \lambda_I, \lambda_D$) remain the same between two updates. Thus for all time $t$ between $t_k$ and $t_{k+1}$, the control signal $\phi(t)$ in Eq.~(\ref{eq:exp-actuator}) equals $\phi(t_k)$.We see that $P$ factor \text{tends} to push the current variable value to the reference value; $I$ factor reduces the accumulative error from the beginning to the current time; $D$ factor controls the fluctuation of the variable.

\subsection{Waterlevel-based Controller}\label{sec:wl}
The Waterlevel-based (WL) controller is another feedback control model which was originally used to switching devices controlled by water level \cite{dezotell1936water}:
\begin{align}
\phi(t_{k+1}) \leftarrow \phi(t_{k}) + \gamma (x_r - x(t_k)), \label{eq:wl}
\end{align}
where $\gamma$ is the step size parameter for $\phi(t_{k})$ update in exponential scale.

Compared to PID, the WL controller only takes the difference between the variable value and the reference value into consideration. Moreover, it provides a sequential control signal. That is, the next control signal is an adjustment based on the previous one.


\begin{figure}
 \centering
 \includegraphics[width=0.98\columnwidth]{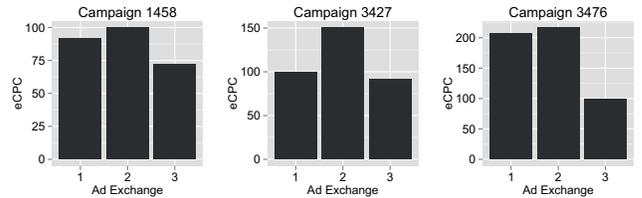}
 \caption{Different eCPCs across different ad exchanges. Dataset: iPinYou.}
 \label{fig:ecpc-adex}
\end{figure}

\subsection{Setting References for Click Maximisation}\label{sec:click-max}
Given that the feedback controller is an effective tool to deliver advertisers' KPI goal, in this subsection, we demonstrate that the feedback control mechanism can be leveraged as a model-free click maximisation framework embedded with any bidding strategies \cite{perlich2012bid,zhang2014optimal} and performs automatic budget allocation \cite{lee2013realtime} across different \emph{channels} via setting smart reference values.


When an advertiser specifies the targeted audience (usually also combined with ad impression contextual categories) for their specific campaign, the impressions that fit the target rules may come from separate channels such as different ad exchanges, user regions, users' PC/mobile devices etc. It is common that the DSP integrates with several ad exchanges and delivers the required ad impressions from all those ad exchanges (as long as the impressions fit the target rule), although the market prices \cite{amin2012budget} may be significantly different. Figure~\ref{fig:ecpc-adex} illustrates that, for the same campaign, there is a difference in terms of eCPC across different ad exchanges.  As pointed out in \cite{zhang2014real}, the differences are also found in other channels such as user regions and devices.

The cost differences provide advertisers a further opportunity to optimise their campaign performance based on eCPCs. To see this, suppose a DSP is integrated to two ad exchanges A and B. For a campaign in this DSP, if its eCPC from exchange A is higher than that from exchange B, which means the inventories from exchange B is more cost effective than those from exchange A, then by reallocating some budget from exchange A to B will \emph{potentially} reduce the overall eCPC of this campaign. Practically the budget reallocation can be done by reducing the bids for exchange A while increasing the bids for exchange B.  Here we formally propose a model of calculating the equilibrium eCPC of each ad exchange, which will be used as the optimal reference eCPC for the feedback control that leads to a maximum number of clicks given the budget constraint.

Mathematically, suppose for a given ad campaign, there are $n$ ad exchanges (could be other channels), i.e., $1, 2, \ldots, n$, that have the ad volume for a target rule. In our formulation we focus on optimising clicks, while the formulation of conversions can be obtained similarly. Let $\xi_i$ be the eCPC on ad exchange $i$, and $c_i(\xi_i)$ be the click number that the campaign acquires in the campaign's lifetime if we tune the bid price to make its eCPC be $\xi_i$ for ad exchange $i$.  For advertisers, they want to maximise the campaign-level click number given the campaign budget $B$ \cite{zhang2014optimal}:

\begin{align}
\max_{\xi_1,\ldots,\xi_n} & \sum_i c_i(\xi_i)\\
\text{s.t.}~~ & \sum_i c_i(\xi_i) \xi_i = B. \label{eq:constraint}
\end{align}

Its Lagrangian is
\begin{align}
\mathcal{L}(\xi_1,\ldots,\xi_n,\alpha) = \sum_i c_i(\xi_i) - \alpha (\sum_i c_i(\xi_i) \xi_i - B), \end{align}
where $\alpha$ is the Lagrangian multiplier. Then we take its gradient on $\xi_i$ and let it be 0:

\begin{align}
\frac{\partial \mathcal{L}(\xi_1,\ldots,\xi_n,\alpha)}{\partial \xi_i} &= c_i'(\xi_i) - \alpha ( c_i'(\xi_i) \xi_i + c_i(\xi_i)) = 0, \label{eq:lag-to-zero}\\
\frac{1}{\alpha} &= \frac{ c_i'(\xi_i) \xi_i + c_i(\xi_i)}{c_i'(\xi_i)} = \xi_i + \frac{c_i(\xi_i)}{c_i'(\xi_i)},
\end{align}
where the equation holds for each ad exchange $i$. As such, we can use $\alpha$ to bridge the equations for any two ad exchanges $i$ and $j$:
\begin{align}
\frac{1}{\alpha} = \xi_i + \frac{c_i(\xi_i)}{c_i'(\xi_i)} = \xi_j + \frac{c_j(\xi_j)}{c_j'(\xi_j)}.
\end{align}

So the optimal solution condition is given as follows:
\begin{align}
\frac{1}{\alpha} = \xi_1 + \frac{c_1(\xi_1)}{c_1'(\xi_1)} = \xi_2 + \frac{c_2(\xi_2)}{c_2'(\xi_2)} &= \cdots = \xi_n + \frac{c_n(\xi_n)}{c_n'(\xi_n)} \label{eq:lagrangian_xi},\\
\sum_i c_i(\xi_i) \xi_i &= B. \label{eq:lambda-condition}
\end{align}

\begin{figure}[t]
 \centering
 \vspace{-5pt}
 \includegraphics[width=0.98\columnwidth]{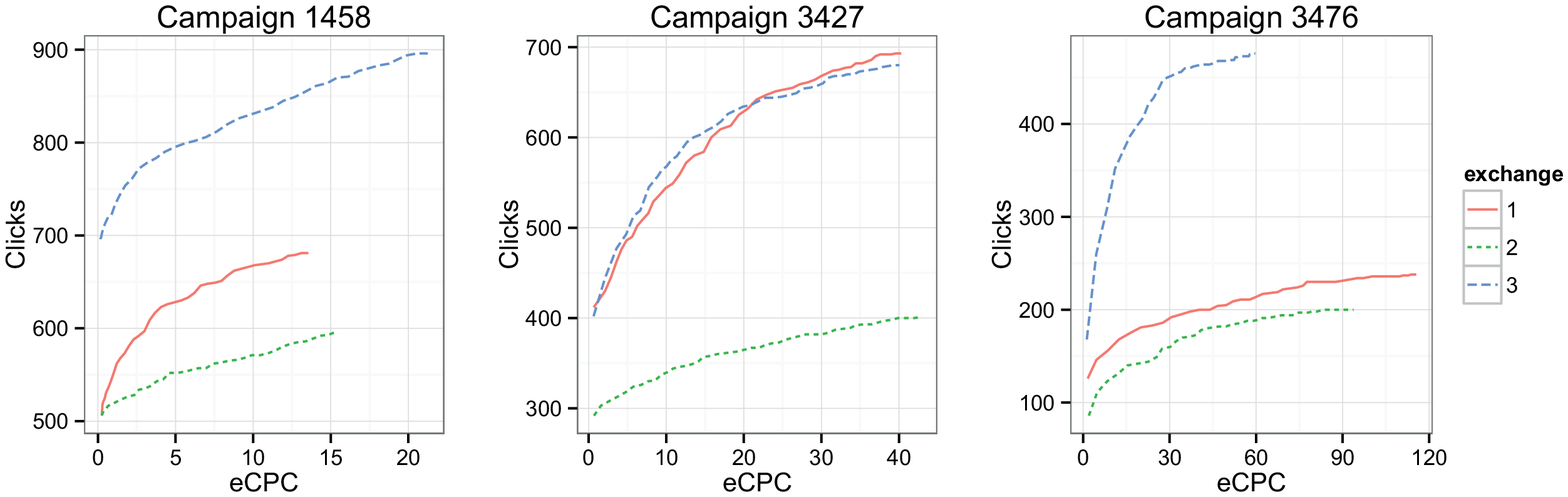}
 \vspace{-5pt}
 \caption{Number of Clicks against eCPC. Clicks and eCPC are calculated across the whole iPinYou training dataset of each campaign by tuning $b_0$ in Eq.~(\ref{eq:bid-lin}).}
 \label{fig:click-ecpc}
\end{figure}

With sufficient data instances, we find that $c_i(\xi_i)$ is usually a concave and smooth function. Some examples are given in Figure~\ref{fig:click-ecpc}. Based on the observation, it is reasonable to define a general polynomial form of the $c_i(\xi_i)$ functions:
\begin{align}
c_i(\xi_i) = c_i^* a_i \Big( \frac{\xi_i}{\xi_i^*}\Big)^{b_i}, \label{eq:click_function}
\end{align}
where $\xi_i^*$ is the campaign's historic average eCPC on the ad inventories from ad exchange $i$ during the training data period, and $c_i^*$ is the corresponding click number.
These two factors are directly obtained from the training data. Parameters $a_i$ and $b_i$ are to be tuned to fit the training data.

Substituting Eq.~(\ref{eq:click_function}) into Eq.~(\ref{eq:lagrangian_xi}) gives
\begin{align}
\frac{1}{\alpha} = \xi_i + \frac{c_i(\xi_i)}{c_i'(\xi_i)} = \xi_i + \frac{\frac{c_i^* a_i}{{\xi_i^*}^{b_i}} \xi_i^{b_i}}{\frac{c_i^* a_i}{{\xi_i^*}^{b_i}} b_i \xi_i^{b_i - 1}} = \Big(1 + \frac{1}{b_i}\Big) \xi_i.
\end{align}
We can then rewrite Eq.~(\ref{eq:lagrangian_xi}) as
\begin{align}
&\frac{1}{\alpha} = \Big(1 + \frac{1}{b_1}\Big) \xi_1 = \Big(1 + \frac{1}{b_2}\Big) \xi_2 = \cdots = \Big(1 + \frac{1}{b_n}\Big) \xi_n \label{eq:lagrangian_xi_clear}.\\
& \text{Thus } \xi_i = \frac{b_i}{\alpha(b_i + 1)}.\label{eq:xi_i_solution}
\end{align}

Interestingly, from Eq.~(\ref{eq:xi_i_solution}) we find that the equilibrium is \emph{not} in the state that the eCPCs from the exchanges are the same. Instead, it is when any amount of budget reallocated among the exchanges does not make any more total clicks; for instance, in a two-exchange case, the equilibrium reaches when the increase of the clicks from one exchange equals the decrease from the other (Eq.~(\ref{eq:lag-to-zero})). More specifically, from Eq.~(\ref{eq:xi_i_solution}) we observe that for ad exchange $i$, if its click function $c_i(\xi_i)$ is quite flat, i.e., the click number increases much slowly as its eCPC increases in a certain area, then its learned $b_i$ should be small. This means the factor $\frac{b_i}{b_i + 1}$ is small as well; then from Eq.~(\ref{eq:xi_i_solution}) we can see the optimal eCPC in ad exchange $i$ should be relatively small.

Substituting Eqs.~(\ref{eq:click_function}) and (\ref{eq:xi_i_solution}) into Eq.~(\ref{eq:constraint}) gives
\begin{align}
\sum_i \frac{c_i^* a_i}{{\xi_i^*}^{b_i}} \Big(\frac{b_i}{b_i + 1} \Big)^{b_i + 1} \Big(\frac{1}{\alpha}\Big)^{b_i + 1} = B, \label{eq:transfered-constaint}
\end{align}
where for simplicity, we denote for each ad exchange $i$, its parameter  $\frac{c_i^* a_i}{{\xi_i^*}^{b_i}} \Big(\frac{b_i}{b_i + 1} \Big)^{b_i + 1}$ as $\delta_i$. This give us a simpler form as:
\begin{align}
\sum_i \delta_i \Big(\frac{1}{\alpha}\Big)^{b_i + 1} = B. \label{eq:transfered-constaint-clear}
\end{align}

There is no closed form to solve Eq.~(\ref{eq:transfered-constaint-clear}) for $\alpha$. However, as $b_i$ cannot be negative and $\sum_i \delta_i (\frac{1}{\alpha})^{b_i + 1}$ monotonically increases against $\frac{1}{\alpha}$, one can easily obtain the solution for $\alpha$ by using a numeric solution such as the stochastic gradient decent or the Newton method \cite{battiti1992first}. Finally, based on the solved $\alpha$, we can find the optimal eCPC $\xi_i$ for each ad exchange $i$ using Eq.~({\ref{eq:xi_i_solution}). In fact, these eCPCs are the reference value we want the campaign to achieve for the corresponding ad exchanges. We can use PID controllers, by setting $x_r$ in \text{Eq.}~(\ref{eq:error-factor}) as $\xi_i$ for each ad exchange $i$, to achieve these reference eCPCs so as to achieve the maximum number of clicks on the campaign level.

As a special case, if we regard the whole volume of the campaign as one channel, this method can be directly used as a general bid optimisation tool. It makes use of the campaign's historic data to decide the optimal eCPC and then the click optimisation is performed by control the eCPC to settle at the optimal eCPC as reference. Note that this multi-channel click maximisation framework is flexible to incorporate any bidding strategies.

\section{Empirical Study}\label{sec:exp}
We conduct comprehensive experiments to study the proposed RTB feedback control mechanism. Our focus in this section is on offline evaluation using a publicly-available real-world dataset. To make our experiment repeatable, we have published the experiment code\footnote{\url{https://github.com/wnzhang/rtbcontrol}}. The online deployment and test on a commercial DSP will be reported in Section \ref{sec:online}.

\subsection{Evaluation Setup}\label{sec:exp-setting}

\textbf{Dataset.} We test our system on a publicly available dataset collected from iPinYou DSP \cite{liao2014ipinyou}. It contains the ad log data from 9 campaigns during 10 days in 2013, which consists of 64.75M bid records, 19.50M impressions, 14.79K clicks and 16K Chinese Yuan (CNY) expense. According to the data publisher \cite{liao2014ipinyou}, the last three-day data of each campaign is split as the test data and the rest as the training data. The dataset disk size is 35GB. More statistics and analysis of the dataset is available in \cite{zhang2014real}. The dataset is in a record-per-row format, where each row consists of three parts: (i) The features for this auction, e.g., the time, location, IP address, the URL/domain of the publisher, ad slot size, user interest segments etc. The features of each record are indexed as a 0.7M-dimension sparse binary vector which is fed into a logistic regression CTR estimator of the bidding strategy in Eq.~(\ref{eq:bid-lin}); (ii) The auction winning price, which is the threshold of the bid to win this auction; (iii) The user feedback on the ad impression, i.e., click or not.

\textbf{Evaluation Protocol.} We follow the evaluation protocol from previous studies on bid optimisation \cite{zhang2014optimal,zhang2014real} and an \text{RTB} contest \cite{liao2014ipinyou} to run our experiment. Specifically, for each data record, we pass the feature information to our bidding agent. Our bidding agent generates a new bid based on the CTR prediction and other parameters in Eq.~(\ref{eq:bid-lin}).  We then compare the generated bid with the logged actual auction winning price. If the bid is higher than the auction winning price, we know the bidding agent has won this auction, paid the winning price, and obtained the ad impression. If from the ad impression record there is a click, then the placement has generated a positive outcome (one click) with a cost equal to the winning price. If there is no click, the placement has resulted in a negative outcome and wasted the money.  The control parameters are updated every 2 hours (as one round).

It is worth mentioning that historical user feedback has been widely used for evaluating information retrieval systems \cite{xu2010improving} and recommender systems \cite{hu2008collaborative}. All of them used historic clicks as a proxy for relevancy to train the prediction model as well as to form the ground truth. Similarly, our evaluation protocol keeps the user contexts, displayed ads (creatives etc.), bid requests, and auction environment unchanged. We intend to answer that under the same context if the advertiser were given a different or better bidding strategy or employed a feedback loop, whether they would be able to get more clicks with the budget limitation. The click would stay the same as nothing has been changed for the users. This methodology works well for evaluating bid optimisation \cite{amin2012budget,zhang2014optimal} and has been adopted in the display advertising industry \cite{liao2014ipinyou}.

\textbf{Evaluation Measures.} We adopt several commonly used measures in feedback control systems \cite{astrom2010feedback}. We define the \emph{error band} as the $\pm 10\%$ interval around the reference value. If the controlled variable settles within this area, we consider that the variable is successfully controlled. The \textbf{speed} of convergence (to the reference value) is also important. Specifically, we evaluate the \emph{rise time} to check how fast the controlled variable will get into the error band. We also use the \emph{settling time} to evaluate how fast the controlled variable will be successfully restricted into the error band.  However, fast convergence may bring the problem of inaccurate control. Thus, two control \textbf{accuracy} measures are introduced. We use the \emph{overshoot} to measure the percentage of value that the controlled variable passes over the reference value. After the settling (called the steady state), we use the \emph{RMSE-SS} to evaluate the root mean square error between the controlled variable value and the reference value. At last, we measure the control \textbf{stability} by calculating the standard deviation of the variable value after the settling, named as \emph{SD-SS}.

For bid optimisation performance, we use the campaign's total achieved click number and eCPC as the prime evaluation measures. We also monitor the impression related performance such as impression number, AWR and CPM.

\textbf{Empirical Study Organisation.} Our empirical study consists of five parts with the focus on controlling two KPIs: eCPC and AWR. (i) In Section~\ref{sec:controllability}, we answer whether the proposed feedback control systems are practically capable of controlling the KPIs. (ii) In Section~\ref{sec:difficulty}, we study the control difficulty with different reference value settings. (iii) In Section~\ref{sec:dynamic-ref}, we focus on the PID controller and investigate its attributes on settling the target variable. (iv) In Section~\ref{sec:bid-opt-exp}, we leverage the PID controllers as a bid optimisation tool and study their performance on optimising the campaign's clicks and eCPC across multiple ad exchanges. (v) Finally, more discussions about PID parameter tuning and online update will be given in Section~\ref{sec:para-tuning}.

\subsection{Control Capability}\label{sec:controllability}

For each campaign, we check the performance of the \text{two} controllers on two KPIs. We first tune the control parameters on the training data to minimise the settling time. Then we adopt the controllers over the test data and observe the performance. The detailed control performance on each campaign is provided in Table~\ref{tab:stage-1-perf-ecpc} for eCPC\footnote{``-" cells mean invalid because of the failure to rise or settle.} and Table~\ref{tab:stage-1-perf-awr} for AWR. Figure~\ref{fig:stage-1} shows the controlled KPI curves against the timesteps (i.e., round). The dashed horizontal line means the reference.

\begin{table}[t]
\center
\caption{Overall control performance on eCPC.}\label{tab:stage-1-perf-ecpc}\vspace{-10pt}
\scriptsize
\begin{tabular}{rrrrrrrr}
Cpg. & Cntr & Rise & Settling & Overshoot & RMSE-SS & SD-SS \\ \hline
\\[-2ex]
\multirow{2}{*}{1458} & PID & 1 & 5 & 7.73 & 0.0325 & 0.0313\\
 & WL & 6 & 36 & 0 & 0.0845 & 0.0103\\
\multirow{2}{*}{2259} & PID & 7 & 7 & 8.03 & 0.0449 & 0.0411\\
 & WL & 6 & - & 0 & - & -\\
\multirow{2}{*}{2261} & PID & 3 & 23 & 17.66 & 0.0299 & 0.0294\\
 & WL & 5 & - & 0 & - & -\\
\multirow{2}{*}{2821} & PID & 17 & 22 & 14.47 & 0.0242 & 0.0216\\
 & WL & - & - & 0 & - & -\\
\multirow{2}{*}{2997} & PID & 17 & 17 & 0.75 & 0.0361 & 0.026\\
 & WL & - & - & 0 & - & -\\
\multirow{2}{*}{3358} & PID & 3 & 7 & 23.89 & 0.0337 & 0.0287\\
 & WL & - & - & 0 & - & -\\
\multirow{2}{*}{3386} & PID & 9 & 13 & 7.90 & 0.0341 & 0.0341\\
 & WL & - & - & 0 & - & -\\
\multirow{2}{*}{3427} & PID & 1 & 12 & 29.03 & 0.0396 & 0.0332\\
 & WL & - & - & 0 & - & -\\
\multirow{2}{*}{3476} & PID & 1 & 5 & 7.64 & 0.0327 & 0.031\\
 & WL & 1 & - & 17.11 & - & -\\
\end{tabular}
\end{table}

\begin{table}[t]
\center
\caption{Overall control performance on AWR.}\label{tab:stage-1-perf-awr}\vspace{-10pt}
\scriptsize
\begin{tabular}{rrrrrrrr}
Cpg. & Cntr & Rise & Settling & Overshoot & RMSE-SS & SD-SS \\ \hline
\\[-2ex]
\multirow{2}{*}{1458} & PID & 4 & 10 & 16.86 & 0.0153 & 0.0093\\
 & WL &  3 & 7 & 0.00 & 0.0448 & 0.0231\\
\multirow{2}{*}{2259} & PID & 4 & 6 & 17.08 & 0.0076 & 0.0072\\
 & WL &  1 & 13 & 3.91 & 0.0833 & 0.0113\\
\multirow{2}{*}{2261} & PID &  1 & 4 & 16.39 & 0.0205 & 0.0203\\
 & WL &  1 & - & 2.02 & - & -\\
\multirow{2}{*}{2821} & PID &  6 & 8 & 16.44 & 0.0086 & 0.0086\\
 & WL &  1 & 3 & 5.77 & 0.0501 & 0.0332\\
\multirow{2}{*}{2997} & PID &  1 & 8 & 13.68 & 0.0151 & 0.0151\\
 & WL &  1 & - & 0.00 & - & -\\
\multirow{2}{*}{3358} & PID &  2 & 8 & 22.08 & 0.0250 & 0.0213\\
 & WL &  1 & 7 & 0.13 & 0.0332 & 0.0211\\
\multirow{2}{*}{3386} & PID &  4 & 8 & 18.85 & 0.0133 & 0.0118\\
 & WL &  1 & 5 & 2.95 & 0.0300 & 0.0291\\
\multirow{2}{*}{3427} & PID &  2 & 6 & 26.63 & 0.0200 & 0.0179\\
 & WL &  3 & 13 & 0.24 & 0.0482 & 0.0257\\
\multirow{2}{*}{3476} & PID &  2 & 6 & 27.15 & 0.0175 & 0.0161\\
 & WL &  1 & 7 & 1.49 & 0.0308 & 0.0271\\
\end{tabular}
\end{table}

\begin{figure}[t]
 \centering
 \vspace{-5pt}
 \includegraphics[width=\columnwidth]{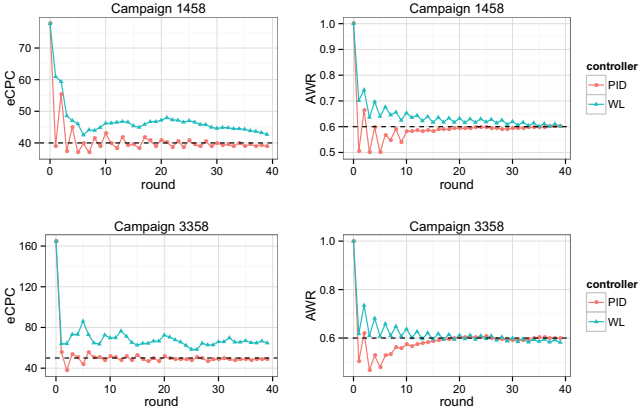}
 \vspace{-16pt}
 \caption{Control performance on eCPC and AWR.}
 \label{fig:stage-1}
\end{figure}

We see from the results that (i) all the PID controllers can settle both KPIs within the error band (with the settling time less than 40 rounds), which indicates that the PID control is capable of settling both KPIs at the given reference value. (ii) The WL controller on eCPC does not work that well on test data, even though we could find good parameters on training data. This is due to the fact that WL controller tries to affect the average system behaviour through transient performance feedbacks while facing the huge dynamics of RTB. (iii) For WL on AWR, most campaigns are controllable while there are still two campaigns that fail to settle at the reference value. (iv) Compared to PID on AWR, WL always results in higher RMSE-SS and SD-SS values but lower overshoot percentage. Those control settings with a fairly short rise time usually face a higher overshoot.
(v) In addition, we observe that the \text{campaigns} with higher \text{CTR} estimator AUC performance (referring \cite{zhang2014real}) normally get shorter settling time.


According to above results, PID controller outperforms the WL controller in the tested RTB cases. We believe this is due to the fact that the integral factor in PID controller helps reduce the accumulative error (i.e., RMSE-SS) and the derivative factor helps reduce the variable fluctuation (i.e., SD-SS). And it is easier to settle the AWR than the eCPC. This is mainly because AWR only depends on the market price distribution while eCPC additionally involves the user feedback, i.e., CTR, where the prediction is associated with significant uncertainty.

\begin{figure}[t]
 \centering
 \subfigure[PID on eCPC]{
 \label{fig:stage-2-pid-ecpc}
 \setcounter{subfigure}{-2}
 	\subfigure{
   		\includegraphics[height=1.1in]{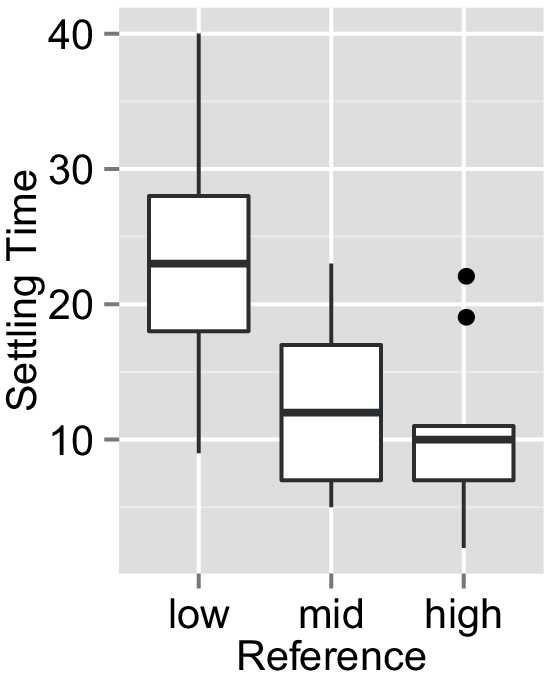}}
   	\subfigure{
   		\includegraphics[height=1.1in]{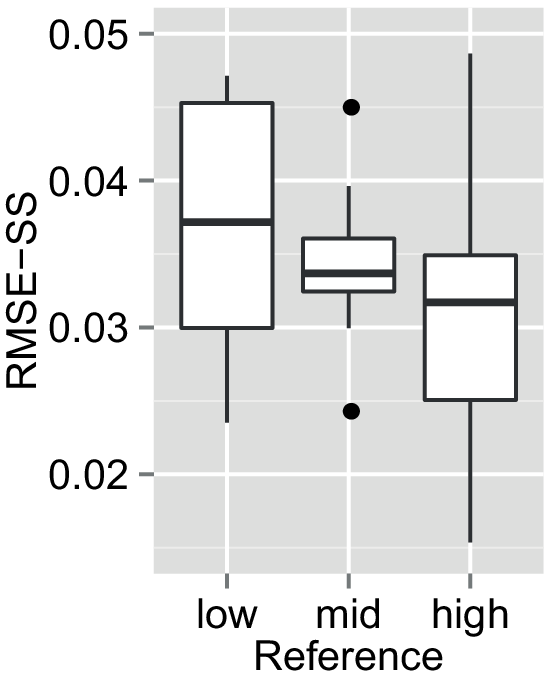}}
   	\subfigure{
	    \includegraphics[height=1.1in]{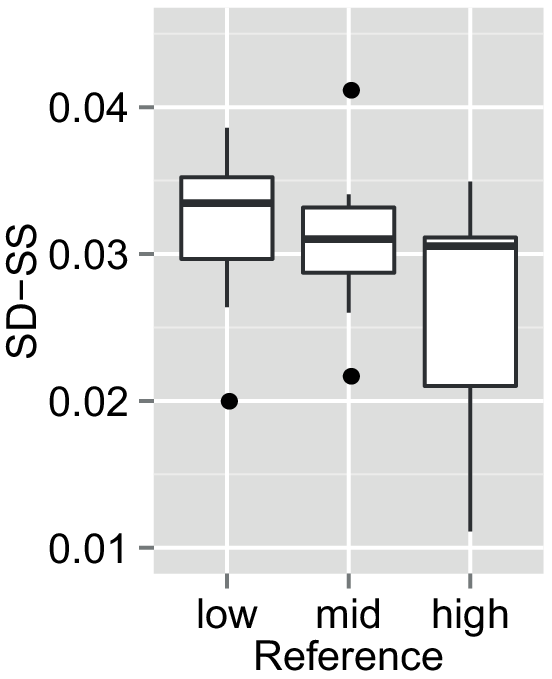}}
   }
 \subfigure[PID on AWR]{
 \label{fig:stage-2-pid-awr}
 \setcounter{subfigure}{-1}
 	\subfigure{
   		\includegraphics[height=1.1in]{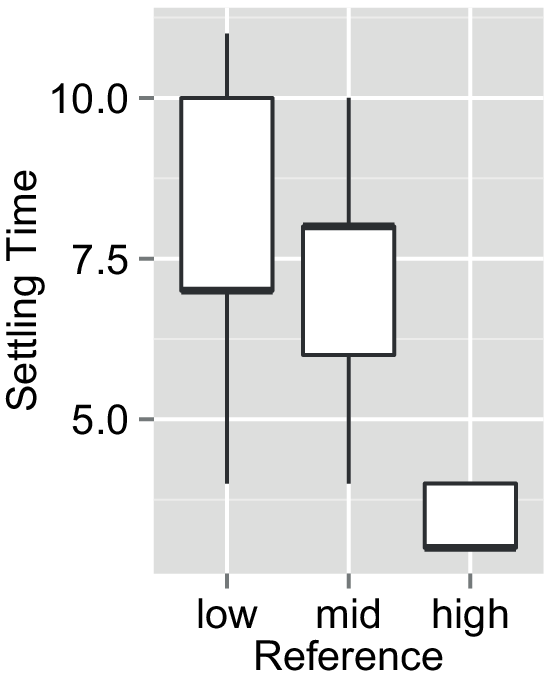}}
   	\subfigure{
   		\includegraphics[height=1.1in]{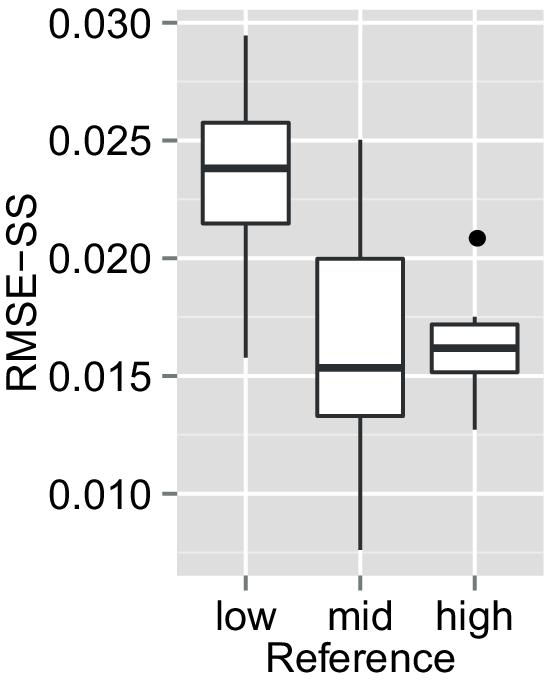}}
   	\subfigure{
	    \includegraphics[height=1.1in]{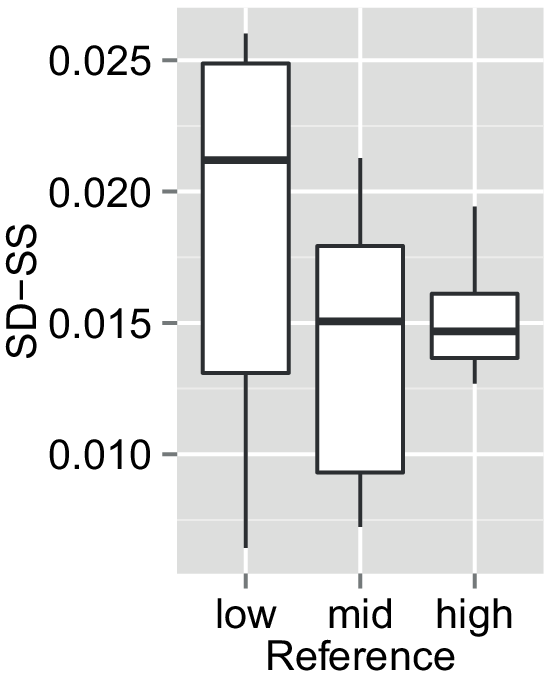}}
   }
   \vspace{-10pt}
 \caption{Control difficulty comparison with PID.}
 \label{fig:stage-2-pid-ecpc-awr}
\end{figure}


\begin{figure}[t]
 \centering
 \vspace{-5pt}
 \includegraphics[width=\columnwidth]{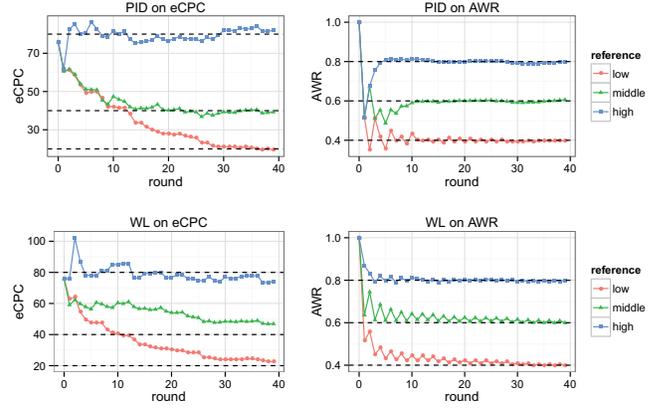}
 \vspace{-10pt}
 \caption{Control performance for campaign 3386 on eCPC and AWR with different reference values.}
 \label{fig:stage-2-3386}
\end{figure}

\subsection{Control Difficulty}\label{sec:difficulty}

In this section, we extend our control capability experiments further by adding higher and lower reference values in comparison. Our goal is to investigate the impact of different levels of reference values on control difficulty. We follow the same scheme to train and test the controllers as Section \ref{sec:controllability}. However, instead of showing the exact performance value, our focus here is on the performance comparison with different reference settings.

The distribution of achieved settling time, RMSE-SS and SD-SS, with the setting of three reference levels, i.e., low, middle and high, are shown in the form of box plot \cite{mcgill1978variations} in the Figure~\ref{fig:stage-2-pid-ecpc} and \ref{fig:stage-2-pid-awr} for the eCPC and AWR control with PID. We observe that the average settling time, RMSE-SS and SD-SS, are reduced as the reference values get higher. This shows that generally the control tasks with higher reference eCPC and AWR are easier to achieve because one can simply bid higher to win more and spend more. Also as the higher reference is closer to the initial performance value, the control signal does not bring serious bias or volatility, which leads to the lower RMSE-SS and SD-SS. For the page limit, the control performance with WL is not presented here. The results are similar with PID.

Figure~\ref{fig:stage-2-3386} gives the specific control curves of the two controllers with three reference levels on a sample campaign 3386. We find that the reference value which is farthest away from the initial value of the controlled variable brings the largest difficulty for settling, both on eCPC and AWR. This suggests that advertisers setting an ambitious control target will introduce the risk of unsettling or large volatility. The advertisers should try to find a best trade-off between the target value and the practical control performance.

\subsection{\mbox{PID Settling: Static vs. Dynamic References}}\label{sec:dynamic-ref}



The combination of proportional, integral and derivative factors enables the PID feedback to automatically adjust the settling progress during the control lifetime with high efficiency \cite{bhattacharyya1995robust}. Alternatively, one can empirically adjust the reference value in order to achieve the desired reference value. For example of eCPC control, if the campaign's achieved eCPC is higher than the initial reference value right after exhausting the first half budget, the advertiser might want to lower the reference value in order to accelerate the downward adjustment and finally reach its initial eCPC target before running out of the budget. PID feedback controller implicitly handles such problem via its integration factor \cite{bhattacharyya1995robust,rivera1986internal}. In this section, we investigate with our RTB feedback control mechanism whether it is still necessary for advertisers to intentionally adjust the reference value according to the campaign's real-time performance.



\textbf{Dynamic Reference Adjustment Model.} To simulate the advertisers' strategies to adaptively change the reference value of eCPC and AWR under the budget constraint, we propose a dynamic reference adjustment model to calculate the new reference $x_r(t_{k+1})$ after $t_k$:
\begin{align} x_r(t_{k+1}) = \frac{(B - s(t_k))x_r x(t_k)}{B x(t_k) - s(t_k) x_r},
\label{eq:general-ref-adjust}
\end{align}
where $x_r$ is the initial reference value, $x(t_k)$ is the achieved KPI (eCPC or AWR) at timestep $t_k$, $B$ is the campaign budget, $s(t_k)$ is the cost so far.  We can see from Eq.~(\ref{eq:general-ref-adjust}) that when $x_r(t_k)=x_r$, $x_r(t_{k+1})$ will be set the same as $x_r$; when $x_r(t_k)>x_r$, $x_r(t_{k+1})$ will be set lower than $x_r$ and vice versa.  For readability, we leave the detailed derivation in appendix. Using Eq.~(\ref{eq:general-ref-adjust}) we calculate the new reference eCPC/AWR $x_r(t_{k+1})$ and use it to substitute $x_r$ in Eq.~(\ref{eq:error-factor}) to calculated the error factor so as to make the dynamic-reference control.  \begin{figure}[t]
 \centering
 \includegraphics[width=\columnwidth]{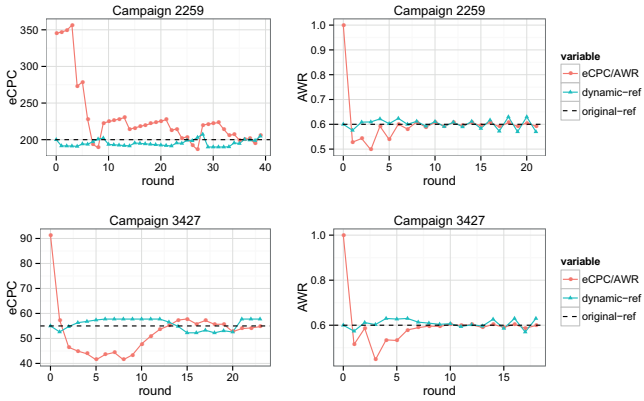}
 \caption{Dynamic reference control with PID.}
 \label{fig:stage-3-dynamic-perf}
\end{figure}

\begin{figure}[t]
 \centering
 \subfigure[PID on eCPC]{
 \label{fig:stage-3-pid-ecpc}
 \setcounter{subfigure}{-3}
 \subfigure{
   \includegraphics[height=1.2in]{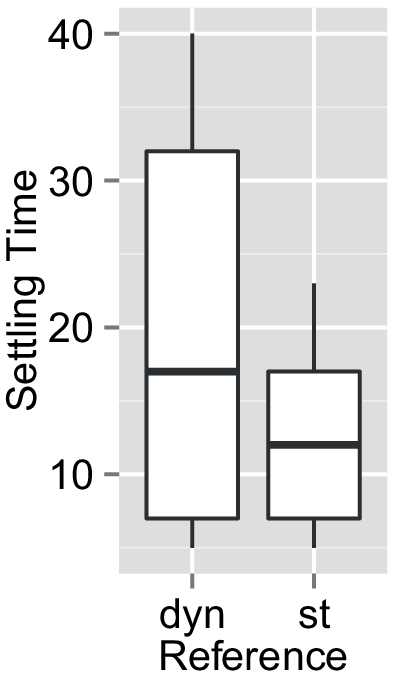}}
 \subfigure{
   \includegraphics[height=1.2in]{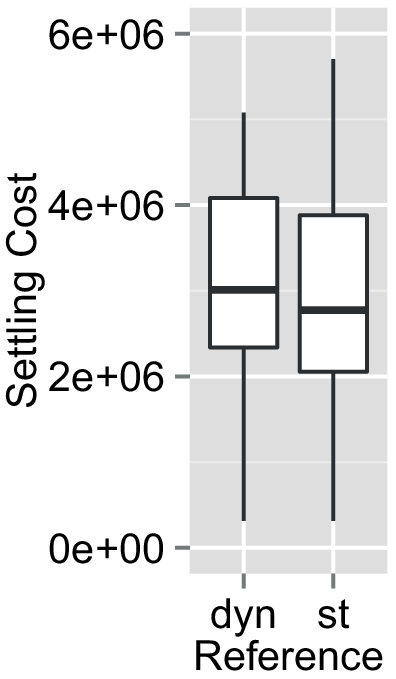}}
 \subfigure{
   \includegraphics[height=1.2in]{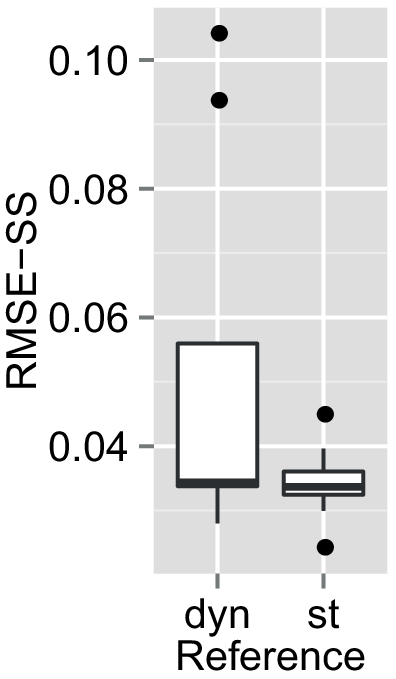}}
 \subfigure{
   \includegraphics[height=1.2in]{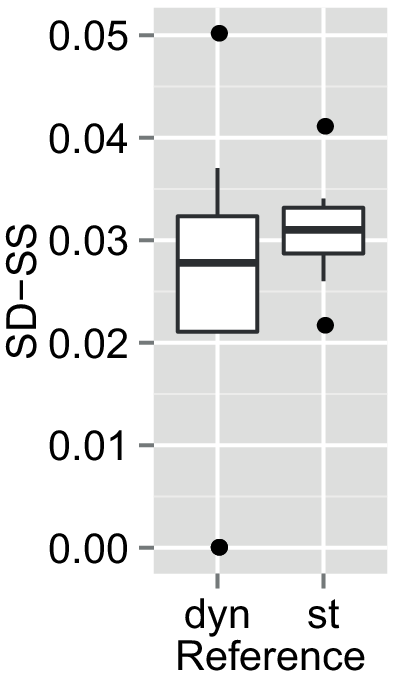}}
   }
 \subfigure[PID on AWR]{
 \label{fig:stage-3-pid-awr}
 \setcounter{subfigure}{-2}
 \subfigure{
   \includegraphics[height=1.2in]{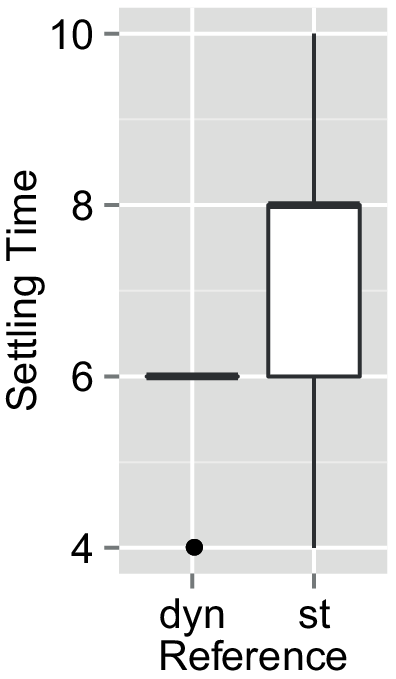}}
 \subfigure{
   \includegraphics[height=1.2in]{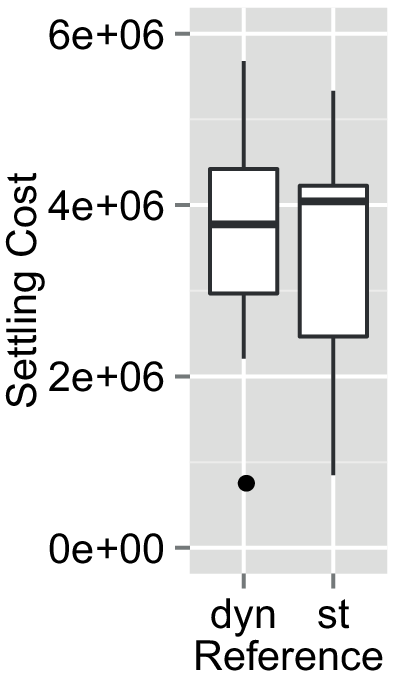}}
 \subfigure{
   \includegraphics[height=1.2in]{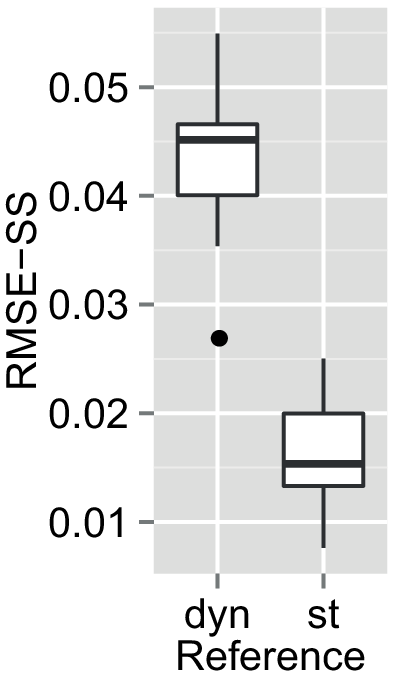}}
 \subfigure{
   \includegraphics[height=1.2in]{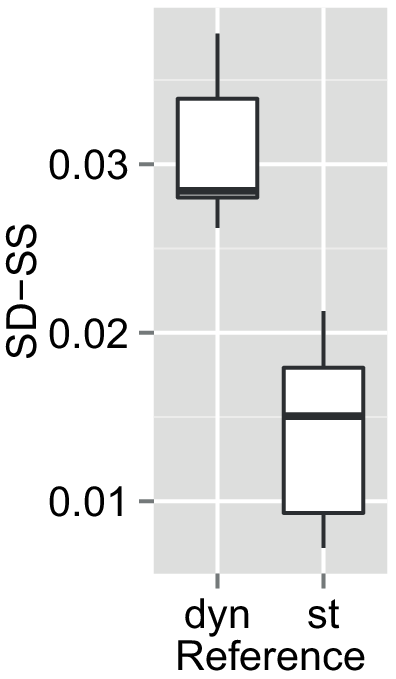}}
 }
 \vspace{-10pt}
\caption{Dynamic v.s. static reference with PID.}
 \label{fig:stage-3-pid}
\end{figure}

\textbf{Results and Discussions.}
Figure~\ref{fig:stage-3-dynamic-perf} shows the PID control performance with dynamic reference calculated based on Eq.~(\ref{eq:general-ref-adjust}). The campaign performance gets stopped at the point where the budget is exhausted. From the figure, we see that for both eCPC and AWR control, the dynamic reference takes an aggressive approach and pushes the eCPC or AWR across the original reference value (dashed line). This actually simulates some advertisers' strategy: when the performance is lower than the reference, then higher the dynamic reference to push the total performance to the initial reference more quickly. Furthermore, for AWR control, we can see the dynamic reference fluctuates seriously when the budget is to be exhausted soon. This is because when there is insufficient budget left, the reference value will be set much high or low by Eq.~(\ref{eq:general-ref-adjust}) in order to push the performance back to the initial target. Apparently this is an ineffective solution.

Furthermore, we directly compare the quantitative control performance between dynamic-reference controllers (\textsf{dyn}) with the standard static-reference ones (\textsf{st}) using PID. Besides the settling time, we also compare the settling cost, which is the spent budget before settling. The overall performance across all the campaigns is shown in Figure~\ref{fig:stage-3-pid-ecpc} for eCPC control and Figure~\ref{fig:stage-3-pid-awr} for AWR control. The results show that (i) for eCPC control, the dynamic-reference controllers do \emph{not} perform better than the static-reference ones; (ii) for AWR control, the dynamic-reference controllers could reduce the settling time and cost, but the accuracy (RMSE-SS) and stability (SD-SS) is much worse than the static-reference controllers. This is because the dynamic reference itself brings volatility (see Figure~\ref{fig:stage-3-dynamic-perf}). These results demonstrate that PID controller does perform a good enough way to settling the variable towards the pre-specified reference without the need of dynamically adjusting the reference to accelerate using our methods. Other dynamic reference models might be somewhat effective but this is not the focus of this paper.

\subsection{Reference Setting for Click Maximisation}\label{sec:bid-opt-exp}
We now study how the proposed feedback control could be used for click optimisation purpose. As we have discussed in Section \ref{sec:click-max}, bid requests usually come from different ad exchanges where the market power and thus the CPM prices are disparate. We have shown that given a budget \text{constraint}, the number of clicks is maximised if one can control the eCPC in each ad exchange by settling it at an optimal eCPC reference for each of them, respectively.


In this experiment, we build a PID feedback controller for each of its integrated ad exchanges, where their reference eCPCs are calculated via Eqs.~(\ref{eq:xi_i_solution}) and (\ref{eq:transfered-constaint-clear}). We train the PID parameters on the training data of each campaign, and then test the bidding performance on the test data. As shown in Table~\ref{tab:stage-4-control-perf-ecpc-adex}, the eCPC on all the ad exchanges for all tested campaigns get settled at the reference values\footnote{Campaign 2997 is only integrated with one ad exchange, thus not compared here.} (settling time less than 40). We denote our multi-exchange eCPC feedback control method as \textsf{multiple}. Besides \textsf{multiple}, we also test a baseline method which assigns a single optimal uniform eCPC reference across all the ad exchanges, denoted as \textsf{uniform}. We also use the linear bidding strategy without feedback control \cite{perlich2012bid} as a baseline, denoted as \textsf{none}\footnote{Other bidding strategies \cite{zhang2014optimal,lee2013realtime} are also investigated. Producing similar results, they are omitted here for clarity.}.


\begin{table}[t]
\center
\caption{Control performance on multi-exchanges with the reference eCPC set for click maximisation.}\label{tab:stage-4-control-perf-ecpc-adex}\vspace{-3pt}
\scriptsize
\begin{tabular}{rrrr|rrrr}
Cpg. & AdEx & Rise & Settling & Cpg. & AdEx & Rise & Settling\\ \hline
\multirow{3}{*}{1458}
& 1 & 13 & 26 & \multirow{3}{*}{3358} & 1 & 9 & 20\\
& 2 & 15 & 18 & & 2 & 14 & 39\\
& 3 & 13 & 13 & & 3 & 26 & 26\\ \hline
\multirow{3}{*}{2259}
& 1 & 10 & 38 & \multirow{3}{*}{3386} & 1 & 6 & 18\\
& 2 & 3 & 14 & & 2 & 12 & 12\\
& 3 & 3 & 29 & & 3 & 1 & 1\\ \hline
\multirow{3}{*}{2261}
& 1 & 3 & 30 & \multirow{3}{*}{3427} & 1 & 16 & 16\\
& 2 & 7 & 38 & & 2 & 35 & 35\\
& 3 & 0 & 35 & & 3 & 23 & 23\\ \hline
\multirow{4}{*}{2821}
& 1 & 6 & 17 & \multirow{4}{*}{3476} & 1 & 18 & 29\\
& 2 & 3 & 10 & & 2 & 22 & 28\\
& 3 & 15 & 15 & & 3 & 19 & 22\\
& 4 & 4 & 38 & & & &
\end{tabular}
\end{table}

\begin{figure}[t]
 \centering
 \includegraphics[width=1\columnwidth]{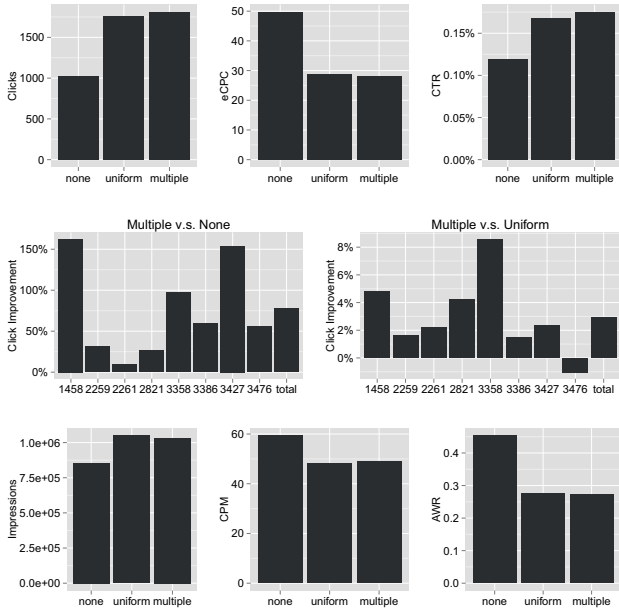}
 \vspace{-15pt}
 \caption{Bid optimisation performance.}\vspace{5pt}
 \label{fig:stage-4-bid-opt-perf}
\end{figure}

The comparisons over various evaluation measures are reported in Figure~\ref{fig:stage-4-bid-opt-perf}. We observe that (i) the feedback-control-enabled bidding strategies \textsf{uniform} and \textsf{multiple} significantly outperform the non-controlled bidding strategy \textsf{none} in terms of the number of achieved clicks and eCPC. This suggests that properly controlling eCPCs would lead to an optimal solution for maximising clicks. (ii) By reallocating the budget via setting different reference eCPCs on different ad exchanges, \textsf{multiple} further outperforms \textsf{uniform} on 7 out of 8 tested campaigns.
(iii) On the impression related measures, the feedback-control-enabled bidding strategies earn more impressions than the non-controlled bidding strategy by actively lowering their bids (CPM) and thus AWR, but achieving more bid volumes. This suggests that by allocating more budget to the lower valued impressions, one could potentially generate more clicks. As a by-product, this confirms the theoretical finding reported in \cite{zhang2014optimal}.

As a case study, Figure~\ref{fig:stage-4-show-case} plots the settling performance of the three methods on campaign 1458. The three dashed horizontal lines are the reference eCPCs on three ad exchanges. We see that the eCPCs on the three ad exchanges successfully settle at the reference eCPCs. At the same time, the campaign-level eCPC (\textsf{multiple}) settles at a lower value than \textsf{uniform} and \textsf{none}.

\begin{figure}[t]
 \centering
 \includegraphics[width=0.8\columnwidth]{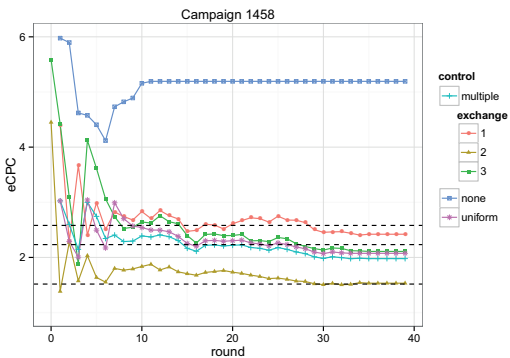}
 \vspace{-5pt}
 \caption{Settlement of multi-exchange feedback control.}
 \label{fig:stage-4-show-case}
\end{figure}

%
%

\subsection{PID Parameter Tuning}\label{sec:para-tuning}
In this subsection, we share some lessons learned about PID controller parameter tuning and online update.

\textbf{Parameter Search.} Empirically, $\lambda_D$ does not change the control performance significantly. Just a small valued $\lambda_D$, e.g., $1\times 10^{-5}$, will reduce the overshoot and slightly shorten the settling time. Thus the parameter search is focused on $\lambda_P$ and $\lambda_I$. Instead of using the computationally expensive grid search, we perform an adaptive coordinate search. For every update, we fix one parameter and shoot another one to seek for the optimal value leading shortest settling time, and the line searching step length shrinks exponentially for each shooting. Normally after 3 or 4 iterations, the local optima is reached and we find such solution is highly comparable with the expensive grid search.

\textbf{Setting $\phi(t)$ Bounds.} We also find that setting up upper/lower bounds of control signal $\phi(t)$ is important to make KPIs controllable. Due to the dynamics in RTB, it is common that user CTR drops during a period, which makes eCPC much higher. The corresponding feedback would probably result in a large negative gain on the bids, leading extremely low bid price and thus no win, no click and no additional cost at all for remaining rounds. In such case, a proper lower bound (-2) of $\phi(t)$ aims to eliminate above extreme effects by preventing from a seriously negative control signal. In addition, an upper bound (5) is used in order to avoid excessive variable growth beyond the reference value.

\textbf{Online Parameter Updating.} As the DSP running with feedback control, the collected data can be immediately utilised for training a new PID controller and updating the older one. We investigate the possibility of the online updating of PID parameters with the recent data. Specifically, after initialising the PID parameters using training data, we re-train the controller for every 10 rounds (i.e., before round 10, 20 and 30) in the test stage using all previous data with the same parameter searching method as in the training stage. The parameter searching in re-training takes about 10 minutes for each controller, which is far shorter than the round period (2 hours). Figure~\ref{fig:online-para-tuning} shows the control performance with PID parameters tuned online and offline respectively. As we can see after the 10th round (i.e., the first online tuning point), the online-tuned PIDs manage to control the eCPC around the reference value more effectively than the offline-tuned one, resulting shorter settling time and lower overshoot. In addition, no obvious disturbance or instability occurs when we switch parameters. With the online parameter updating, we can start to train the controllers based on several-hour training data and adaptively update the parameters from the new data to improve the control performance.

\begin{figure}[t]
 \centering
 \includegraphics[width=1\columnwidth]{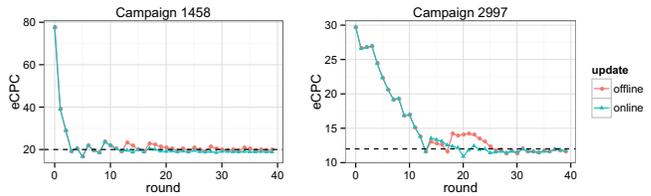}
 \vspace{-15pt}
 \caption{Control with online/offline parameter updating.}
 \label{fig:online-para-tuning}
\end{figure}

\section{Online Deployment and Test}\label{sec:online}
The proposed RTB feedback control system has been deployed and tested in live on BigTree DSP\footnote{\url{http://www.bigtree.mobi/}}, a performance-driven mobile advertising DSP in China. BigTree DSP focuses on the programmatic buying for optimal advertising performance on mobile devices, which makes it an ideal place to test our proposed solution.

The deployment environment is based on Aliyun elastic cloud computing servers. A three-node cluster is deployed for the DSP bidding agent, where each node is in Ubuntu 12.04, with 8 core Intel Xeon CPU E5-2630 (2.30GHz) and 8GB RAM. The controller module is implemented in Python with uWSGI and Nginx.

For BigTree DSP controller module, we deploy the PID control function and tune its parameters. Specifically, we use the last 6-week bidding log data in 2014 as the training data for tuning PID parameters. A three-fold validation process is performed to evaluate the generalisation of the PID control performance, where the previous week data is used as the training data while the later week data is used for validation. The control factors ($\phi(t), e(t_k)$ in Eq.~(\ref{eq:pid})) are updated for every 90 minutes. After acquiring a set of robust and effective PID parameters, we launch the controller module, including the monitor and actuator submodules, on BigTree DSP.

\begin{figure}[t]
 \centering
 \includegraphics[width=0.95\columnwidth]{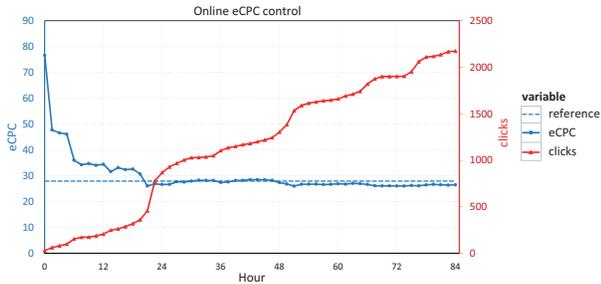}\vspace{-5pt}
 \caption{The online eCPC control performance and the accumulative click numbers of a mobile game campaign on BigTree DSP.} 
 \label{fig:stage-online-ecpc}
\end{figure}

\begin{figure}[t]
 \vspace{-5pt}
 \centering
 \includegraphics[width=0.98\columnwidth]{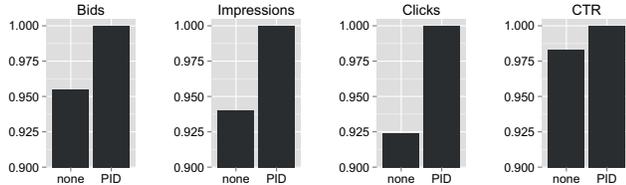}\vspace{-5pt}
 \caption{Relative performance for online test.}
 \label{fig:stage-online-performance}
\end{figure}

Figure~\ref{fig:stage-online-ecpc} shows the online eCPC control performance on one of the iOS mobile game campaigns during 84 hours from 7 Jan. 2015 to 10 Jan. 2015. The reference eCPC is set as 28 RMB cent by the advertiser, which is about 0.8 times the average eCPC value of the previous week where there was no control. Following the same training process described in the previous section, we update the online control factors for every 90 minutes. From the result we can see the eCPC value dropped from the beginning 79 cent to 30 during the first day and then settled closed to the reference afterwards.

In the meantime, A/B testing is used to compare with the non-controlled bidding agent (with the same sampling rate but disjoint bid requests). Figure~\ref{fig:stage-online-performance} shows the corresponding advertising performance comparison between a non-controlled bidding agent and the PID-control bidding agent during the test period with the same budget. As we can see, by settling the eCPC value around the lower reference eCPC, the PID-control bidding agent acquires more bid volume and win more (higher-CTR) impressions and \text{clicks}, which demonstrates its ability of optimising the performance.

Compared with the offline empirical study, the online running is more challenging: (i) all pipeline steps including the update of the CTR estimator, the KPI monitor linked to the database and the PID controller should operate smoothly against the market turbulence; (ii) the real market competition is highly dynamic during the new year period when we launched our test; (iii) other competitors might tune their bidding strategies independently or according to any changes of their performance after we employed the controlled bidding strategy. In sum, the successful eCPC control on an online commercial DSP verifies the effectiveness of our proposed feedback control RTB system.

\vspace{-5pt}
\section{Related Work}\label{sec:related-work}


Enabling the impression-level evaluation and bidding, much research work has been done on RTB display advertising, including bidding strategy optimisation \cite{perlich2012bid,zhang2014optimal}, reserve price optimisation \cite{yuan2014empirical}, ad exchange auction design \cite{balseiro2014repeated}, and ad tracking \cite{gomer2013network}.

In order to perform the optimal bidding, the DSP bidding agent should estimate both utility and cost of a given ad impression. The impression-level utility evaluation, including CTR and conversion rate (CVR) estimation, is the essential part for each bidding agent in DSPs. In \cite{lee2012estimating} the sparsity problem of CVR estimation is handled by modelling the conversions at different hierarchical levels. The user click behaviour on mobile RTB ads is studied in \cite{oentaryo2014predicting}. On the cost evaluation side, bid landscape modelling and forecasting is much important to inform the bidding agent about the competitiveness of the market. The authors in \cite{cui2011bid} break down the campaign-level bid landscape forecasting problem into ``samples'' by targeting rules and then employ a mixture model of log-normal distributions to build the campaign-level bid landscape. The authors in \cite{lang2012handling} try to reduce the bid landscape forecasting error through frequently re-building the landscape models. Based on the utility and cost evaluation of the ad inventory, bid optimisation is performed to improve the advertising performance under the campaign budget constraint. Given the estimated CTR/CVR, the authors in \cite{perlich2012bid,lee2012estimating} employ linear bidding functions based on truth-telling attributes of second price auctions. However, given the budget constraint, the advertisers' bidding behaviour is not truth-telling. The authors in \cite{zhang2014optimal} propose a general bid optimisation framework to maximise the desired advertising KPI (e.g., total click number) under the budget constraint. Besides the general bid optimisation, the explicit bidding rules such as frequency and recency capping are studied in \cite{yuan2013real}. Moreover, the budget pacing \cite{lee2013realtime} which refers to smoothly delivering the campaign budget is another important problem for DSPs.

There are a few research papers on recommender \text{systems} leveraging feedback controllers for performance maintenance and improvement. In \cite{meng2006control}, a rating updating algorithm based on the PID controller is developed to exclude unfair ratings in order to build a robust reputation system. The authors in \cite{zanardi2011dynamic} apply a self-monitoring and self-adaptive approach to perform a dynamic update of the training data fed into the recommender system to automatically balance the computational cost and the prediction accuracy. Furthermore, the authors in \cite{jambor2012using} adopt the more effective and well-studied PID controller to the data-feeding scheme of recommender systems, which is proved to be practically effective in their studied training task.

Compared to the work of controlling the recommender system performance by changing the number of training cases, our control task in RTB is more challenging, with much various dynamics from advertising environment such as the fluctuation of market price, auction volume and user behaviour. In \cite{Chen2011c}, the authors discuss multiple aspects in a performance-driven RTB system, where the impression volume control is one of discussed aspects. Specifically, WL and a model-based controller are implemented to control the impression volume during each time interval. In \cite{karlsson2013applications}, feedback control is used to perform budget pacing in order to stablise the conversion volume. Compared to \cite{Chen2011c,karlsson2013applications}, our work is a more comprehensive study focused on the feedback control techniques to address the instability problem in RTB. Besides WL, we intensively investigate the PID controller, which takes more factors into consideration than WL. For the controlled KPIs, we look into the control tasks on both eCPC and AWR, which are crucial KPIs for performance-driven campaigns and branding-based campaigns, respectively. In addition, we proposed an effective model to calculate the optimal eCPC reference to maximise the campaign's clicks using feedback controllers.



\section{Conclusions}\label{sec:conclusion}
In this paper, we have proposed a feedback control mechanism for RTB display advertising, with the aim of improving its robustness of achieving the advertiser's KPI goal. We mainly studied PID and WL controllers for controlling the eCPC and AWR KPIs. Through our comprehensive empirical study, we have the following discoveries. (i) Despite of the high dynamics in RTB, the KPI variables are controllable using our feedback control mechanism. (ii) Different reference values bring different control difficulties, which are reflected in the control speed, accuracy and stability. (iii) PID controller naturally finds its best way to settle the variable, and there is no necessity to adjust the reference value for accelerating the PID settling. (iv) By settling the eCPCs to the optimised reference values, the feedback controller is capable of making bid optimisation. Deployed on a commercial DSP, the online test demonstrates the effectiveness of the feedback control mechanism in generating controllable advertising performance. In the future work, we will further study the applications based on feedback controllers in RTB, such as budget pacing and retargeting frequency capping.


\vspace{0pt}
{\scriptsize
\bibliographystyle{abbrv}
\bibliography{wsdm560-zhang}
}

\appendix

{

\textbf{Reference Adjust Models.} Here we provide the detailed derivation of the proposed dynamic-reference model Eq.~(\ref{eq:general-ref-adjust}) in Section~\ref{sec:dynamic-ref}.
We mainly introduce the derivation of reference eCPC adjustment, while the derivation of reference AWR adjustment can be obtained similarly.

Let $\xi_r$ be the initial eCPC target, $\xi(t_k)$ be the achieved eCPC before the moment $t_k$, $s(t_k)$ be the total cost so far, and $B$ be the campaign budget. In such setting, the current achieved click number is $s(t_k)/\xi(t_k)$ and the target click number is $B/\xi_r$. In order to push the overall eCPC to $\xi_r$, i.e., push the total click number $B/\xi_r$ with the budget $B$, the reference eCPC for the remaining time $\xi_r(t_{k+1})$ should satisfy
\begin{align}
\frac{s(t_k)}{\xi(t_k)} + \frac{B - s(t_k)}{\xi_r(t_{k+1})} = \frac{B}{\xi_r}.\label{eq:ecpc}
\end{align}
Solving the equation we have
\begin{align}
\xi_r(t_{k+1}) = \frac{(B - s(t_k))\xi_r \xi(t_k)}{B \xi(t_k) - s(t_k) \xi_r}.\label{eq:dynamic-ecpc}
\end{align}

The derivation of reference AWR adjustment is much similar but with an extra winning function which links between the bid price and winning probability \cite{zhang2014optimal}. The result formula is just the same as Eq.~(\ref{eq:dynamic-ecpc}). Using $x$ as a general notation for eCPC and AWR variables results in Eq.~(\ref{eq:general-ref-adjust}) in Section~\ref{sec:dynamic-ref}.

}
\end{document}